\pdfoutput=1

\documentclass[12pt,a4paper]{article}

\usepackage{ifthen} 
\newboolean{pdflatex}
\setboolean{pdflatex}{true} 

\newboolean{articletitles}
\setboolean{articletitles}{true} 

\newboolean{uprightparticles}
\setboolean{uprightparticles}{false} 

\def\paperauthors{LHCb collaboration} 
\def\paperasciititle{Measurement of CP-violating and mixing-induced observables in Bs0 to phi gamma decays} 
\def\papertitle{Measurement of $CP$-violating and mixing-induced observables in $B_s^0 \to \phi\gamma$ decays} 
\def\paperkeywords{{High Energy Physics}, {LHCb}} 
\def\papercopyright{\the\year\ CERN for the benefit of the LHCb collaboration} 
\def\paperlicence{CC-BY-4.0 licence}
\def\paperlicenceurl{https://creativecommons.org/licenses/by/4.0/}

\usepackage[top=1in, bottom=1.25in, left=1in, right=1in]{geometry}

\usepackage{microtype}
\usepackage{lineno}  
\usepackage{xspace} 
\usepackage{caption} 

\usepackage{graphicx}  
\usepackage{color}
\usepackage{colortbl}
\graphicspath{{./figs/}} 
\DeclareGraphicsExtensions{.pdf,.PDF,png,.PNG}

\usepackage{amsmath} 
\usepackage{amssymb}
\usepackage{amsfonts}
\usepackage{upgreek} 

\newcommand*\patchAmsMathEnvironmentForLineno[1]{%
\expandafter\let\csname old#1\expandafter\endcsname\csname #1\endcsname
\expandafter\let\csname oldend#1\expandafter\endcsname\csname
end#1\endcsname
 \renewenvironment{#1}%
   {\linenomath\csname old#1\endcsname}%
   {\csname oldend#1\endcsname\endlinenomath}%
}
\newcommand*\patchBothAmsMathEnvironmentsForLineno[1]{%
  \patchAmsMathEnvironmentForLineno{#1}%
  \patchAmsMathEnvironmentForLineno{#1*}%
}
\AtBeginDocument{%
\patchBothAmsMathEnvironmentsForLineno{equation}%
\patchBothAmsMathEnvironmentsForLineno{align}%
\patchBothAmsMathEnvironmentsForLineno{flalign}%
\patchBothAmsMathEnvironmentsForLineno{alignat}%
\patchBothAmsMathEnvironmentsForLineno{gather}%
\patchBothAmsMathEnvironmentsForLineno{multline}%
\patchBothAmsMathEnvironmentsForLineno{eqnarray}%
}

\usepackage{hyperxmp}

\usepackage[pdftex,
            pdfauthor={\paperauthors},
            pdftitle={\paperasciititle},
            pdfkeywords={\paperkeywords},
            pdfcopyright={Copyright (C) \papercopyright},
            pdflicenseurl={\paperlicenceurl}]{hyperref}

\usepackage[colorinlistoftodos,textsize=scriptsize]{todonotes}

\usepackage[all]{hypcap} 

\usepackage{xspace} 
\usepackage{upgreek}

\newcommand{\offsetoverline}[2][0.1em]{\kern #1\overline{\kern -#1 #2}}%

\def\lhcb   {\mbox{LHCb}\xspace}

\def\MagUp {\mbox{\em Mag\kern -0.05em Up}\xspace}

\ifthenelse{\boolean{uprightparticles}}%
{

 \def\Pmu         {\ensuremath{\upmu}\xspace}

 \def\Ppi         {\ensuremath{\uppi}\xspace}

 \def\Ppsi        {\ensuremath{\uppsi}\xspace}

 \def\PDelta      {\ensuremath{\Delta}\xspace}                 
 \def\PXi         {\ensuremath{\Xi}\xspace}                 
 \def\PLambda     {\ensuremath{\Lambda}\xspace}                 
 \def\PSigma      {\ensuremath{\Sigma}\xspace}                 
 \def\POmega      {\ensuremath{\Omega}\xspace}                 
 \def\PUpsilon    {\ensuremath{\Upsilon}\xspace}

 \def\PB      {\ensuremath{\mathrm{B}}\xspace}                 
                  
 \def\PD      {\ensuremath{\mathrm{D}}\xspace}

 \def\PJ      {\ensuremath{\mathrm{J}}\xspace}                 
 \def\PK      {\ensuremath{\mathrm{K}}\xspace}

 \def\Pb      {\ensuremath{\mathrm{b}}\xspace}

 \def\Pi      {\ensuremath{\mathrm{i}}\xspace}

 \def\Ps      {\ensuremath{\mathrm{s}}\xspace}

}
{

 \def\Pmu         {\ensuremath{\mu}\xspace}

 \def\Ppi         {\ensuremath{\pi}\xspace}

 \def\Ppsi        {\ensuremath{\psi}\xspace}                 
                  
 \mathchardef\PDelta="7101
 \mathchardef\PXi="7104
 \mathchardef\PLambda="7103
 \mathchardef\PSigma="7106
 \mathchardef\POmega="710A
 \mathchardef\PUpsilon="7107
                  
 \def\PB      {\ensuremath{B}\xspace}                 
                  
 \def\PD      {\ensuremath{D}\xspace}

 \def\PJ      {\ensuremath{J}\xspace}                 
 \def\PK      {\ensuremath{K}\xspace}

 \def\Pb      {\ensuremath{b}\xspace}

 \def\Pi      {\ensuremath{i}\xspace}

 \def\Ps      {\ensuremath{s}\xspace}

}

\makeatletter
\ifcase \@ptsize \relax
  \newcommand{\miniscule}{\@setfontsize\miniscule{4}{5}}
\or
  \newcommand{\miniscule}{\@setfontsize\miniscule{5}{6}}
\or
  \newcommand{\miniscule}{\@setfontsize\miniscule{5}{6}}
\fi
\makeatother

\DeclareRobustCommand{\optbar}[1]{\shortstack{{\miniscule (\rule[.5ex]{1.25em}{.18mm})}
  \\ [-.7ex] $#1$}}

\def\mup        {{\ensuremath{\Pmu^+}}\xspace}
\def\mun        {{\ensuremath{\Pmu^-}}\xspace} 

\def\squark    {{\ensuremath{\Ps}}\xspace}

\def\bquark    {{\ensuremath{\Pb}}\xspace}

\def\pion   {{\ensuremath{\Ppi}}\xspace}
\def\piz    {{\ensuremath{\pion^0}}\xspace}

\def\pim    {{\ensuremath{\pion^-}}\xspace}

\def\kaon    {{\ensuremath{\PK}}\xspace}
  \def\Kbar    {{\kern 0.2em\overline{\kern -0.2em \PK}{}}\xspace}

\def\KorKbar {\kern 0.18em\optbar{\kern -0.18em K}{}\xspace}

\def\Kp      {{\ensuremath{\kaon^+}}\xspace}
\def\Km      {{\ensuremath{\kaon^-}}\xspace}

\def\Kstarz  {{\ensuremath{\kaon^{*0}}}\xspace}

  \def\Dbar    {{\kern 0.2em\overline{\kern -0.2em \PD}{}}\xspace}

\def\DorDbar {\kern 0.18em\optbar{\kern -0.18em D}{}\xspace}

\def\Dzb     {{\ensuremath{\Dbar{}^0}}\xspace}

\def\B       {{\ensuremath{\PB}}\xspace}
\def\Bbar    {{\ensuremath{\kern 0.18em\overline{\kern -0.18em \PB}{}}}\xspace}

\def\BorBbar    {\kern 0.18em\optbar{\kern -0.18em B}{}\xspace}

\def\Bd      {{\ensuremath{\B^0}}\xspace}
\def\Bs      {{\ensuremath{\B^0_\squark}}\xspace}
\def\Bsb     {{\ensuremath{\Bbar{}^0_\squark}}\xspace}

\def\jpsi     {{\ensuremath{{\PJ\mskip -3mu/\mskip -2mu\Ppsi\mskip 2mu}}}\xspace}

\def\Y#1S{\ensuremath{\PUpsilon{(#1S)}}\xspace}

\def\Lz          {{\ensuremath{\PLambda}}\xspace}

\def\LorLbar     {\kern 0.18em\optbar{\kern -0.18em \PLambda}{}\xspace}

\def\Lb           {{\ensuremath{\Lz^0_\bquark}}\xspace}

\def\to                 {\ensuremath{\rightarrow}\xspace}

\def\CP                {{\ensuremath{C\!P}}\xspace}

\def\AT#1     {\ensuremath{A_{\mathrm{T}}^{#1}}\xspace}           

\def\C#1      {\ensuremath{\mathcal{C}_{#1}}\xspace}                       
\def\Cp#1     {\ensuremath{\mathcal{C}_{#1}^{'}}\xspace}                    
\def\Ceff#1   {\ensuremath{\mathcal{C}_{#1}^{\mathrm{(eff)}}}\xspace}        
\def\Cpeff#1  {\ensuremath{\mathcal{C}_{#1}^{'\mathrm{(eff)}}}\xspace}       
\def\Ope#1    {\ensuremath{\mathcal{O}_{#1}}\xspace}                       
\def\Opep#1   {\ensuremath{\mathcal{O}_{#1}^{'}}\xspace}                    

\newcommand{\aunit}[1]{\ensuremath{\text{\,#1}}}       

\newcommand{\tev}{\aunit{Te\kern -0.1em V}\xspace}
\newcommand{\gev}{\aunit{Ge\kern -0.1em V}\xspace}
\newcommand{\mev}{\aunit{Me\kern -0.1em V}\xspace}
\newcommand{\kev}{\aunit{ke\kern -0.1em V}\xspace}
\newcommand{\ev}{\aunit{e\kern -0.1em V}\xspace}
\newcommand{\mevc}{\ensuremath{\aunit{Me\kern -0.1em V\!/}c}\xspace}
\newcommand{\gevc}{\ensuremath{\aunit{Ge\kern -0.1em V\!/}c}\xspace}
\newcommand{\mevcc}{\ensuremath{\aunit{Me\kern -0.1em V\!/}c^2}\xspace}
\newcommand{\gevcc}{\ensuremath{\aunit{Ge\kern -0.1em V\!/}c^2}\xspace}

\def\fb   {\ensuremath{\aunit{fb}}\xspace}
\def\invfb   {\ensuremath{\fb^{-1}}\xspace}

\def\ps   {\ensuremath{\aunit{ps}}\xspace}
\def\fs   {\aunit{fs}}

\def\invps{\ensuremath{\ps^{-1}}\xspace}

\def\gsim{{~\raise.15em\hbox{$>$}\kern-.85em
          \lower.35em\hbox{$\sim$}~}\xspace}
\def\lsim{{~\raise.15em\hbox{$<$}\kern-.85em
          \lower.35em\hbox{$\sim$}~}\xspace}

\def\PDF {PDF\xspace}

\def\pt         {\ensuremath{p_{\mathrm{T}}}\xspace}

\def\ptot       {\ensuremath{p}\xspace}

\xspace

\def\tell1  {TELL1\xspace}
\def\ukl1   {UKL1\xspace}

\usepackage{cite} 
\usepackage{mciteplus}

\newcommand{\phigamma}{\ensuremath{B_s^0 \to \phi\gamma}\xspace}
\newcommand{\kstgamma}{\ensuremath{B^0 \to \Kstarz\gamma}\xspace}

\newcommand{\adeltavanilla}{\ensuremath{\mathcal{A}^{\Delta}}\xspace}
\newcommand{\Svanilla}{\ensuremath{S}\xspace}
\newcommand{\Cvanilla}{\ensuremath{C}\xspace}
\newcommand{\adelta}{\ensuremath{\mathcal{A}^{\Delta}_{\phi\gamma}}\xspace}
\newcommand{\Spar}{\ensuremath{S_{\phi\gamma}}\xspace}
\newcommand{\Cpar}{\ensuremath{C_{\phi\gamma}}\xspace}

\begin{document}

\renewcommand{\thefootnote}{\fnsymbol{footnote}}
\setcounter{footnote}{1}


\begin{titlepage}
\pagenumbering{roman}

\vspace*{-1.5cm}
\centerline{\large EUROPEAN ORGANIZATION FOR NUCLEAR RESEARCH (CERN)}
\vspace*{1.5cm}
\noindent
\begin{tabular*}{\linewidth}{lc@{\extracolsep{\fill}}r@{\extracolsep{0pt}}}
\ifthenelse{\boolean{pdflatex}}
{\vspace*{-1.5cm}\mbox{\!\!\!\includegraphics[width=.14\textwidth]{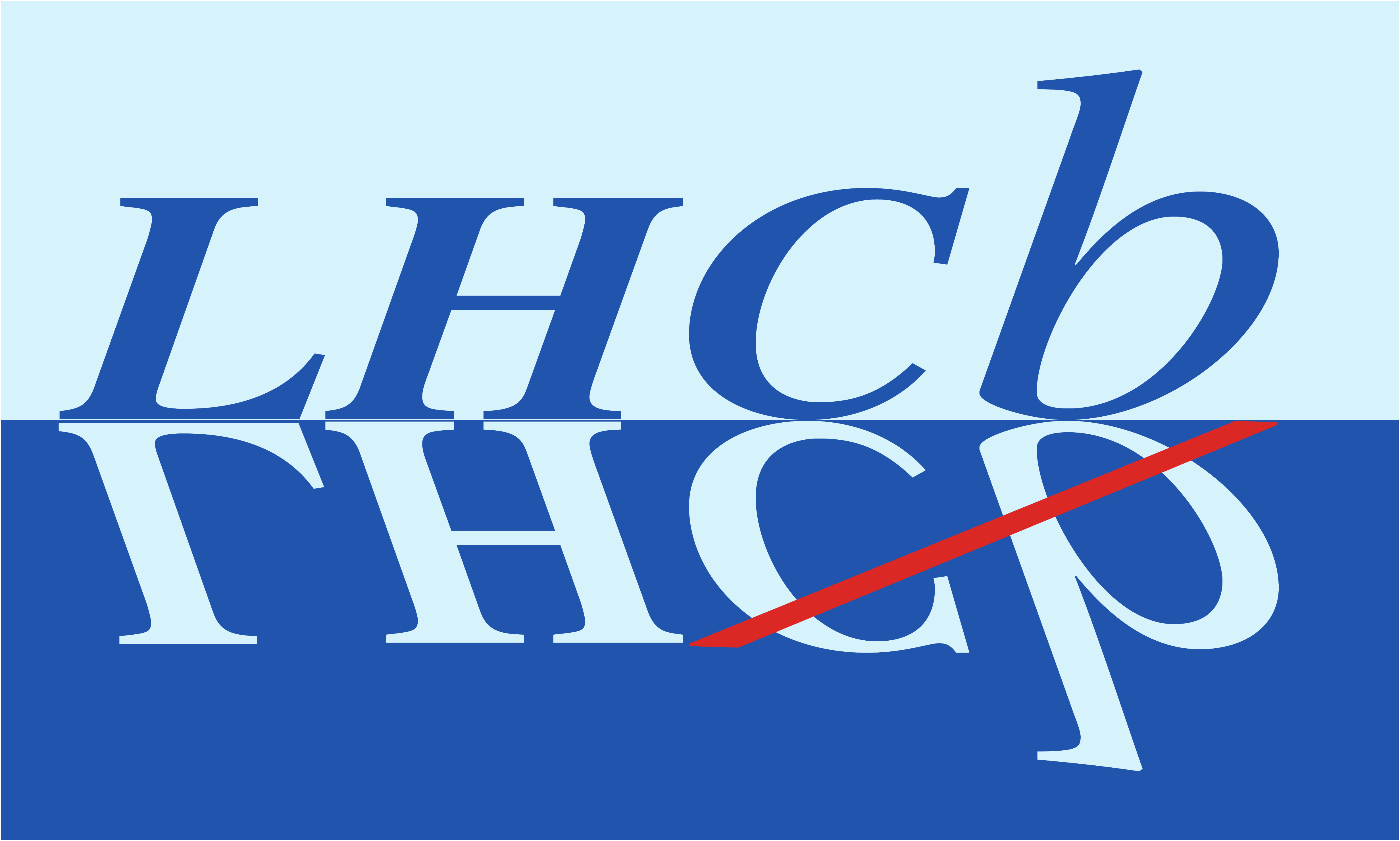}} & &}%
{\vspace*{-1.2cm}\mbox{\!\!\!\includegraphics[width=.12\textwidth]{figs/lhcb-logo.eps}} & &}%
\\
 & & CERN-EP-2019-077 \\  
 & & LHCb-PAPER-2019-015 \\  
 & & August 29, 2019 \\ 
 & & \\
\end{tabular*}

\vspace*{3.0cm}

{\normalfont\bfseries\boldmath\huge
\begin{center}
  \papertitle 
\end{center}
}

\vspace*{1.0cm}

\begin{center}
\paperauthors\footnote{Authors are listed at the end of this Letter.}
\end{center}

\vspace{\fill}

\begin{abstract}
  \noindent
A time-dependent analysis of the $B_s^0 \to \phi\gamma$ decay rate is performed to determine the $CP$-violating observables $S_{\phi\gamma}$ and $C_{\phi\gamma}$, and the mixing-induced observable $\mathcal{A}^{\Delta}_{\phi\gamma}$.
The measurement is based on a sample of $pp$ collision data recorded with the LHCb detector, corresponding to an integrated luminosity of 3 fb$^{-1}$ at center-of-mass energies of 7 and 8 TeV.
  The measured values are
  \begin{align*}
  	S_{\phi\gamma} &= 0.43 \pm 0.30 \pm 0.11, \\
  	C_{\phi\gamma} &= 0.11 \pm 0.29 \pm 0.11, \\
  	\mathcal{A}^{\Delta}_{\phi\gamma} &= -0.67 \, ^{+0.37}_{-0.41} \pm 0.17,
  \end{align*}
  where the first uncertainty is statistical and the second systematic.
  This is the first measurement of the observables $S$ and $C$ in radiative $B_s^0$ decays.
  The results are consistent with the Standard Model predictions.
\end{abstract}

\vspace{\fill}

\begin{center}
  Published in Phys.~Rev.~Lett. 123 (2019) 081802
\end{center}

\vspace{\fill}

{\footnotesize 
\centerline{\copyright~\papercopyright. \href{\paperlicenceurl}{\paperlicence}.}}
\vspace*{2mm}

\end{titlepage}


\newpage
\setcounter{page}{2}
\mbox{~}

\cleardoublepage

\renewcommand{\thefootnote}{\arabic{footnote}}
\setcounter{footnote}{0}


\pagestyle{plain}
\setcounter{page}{1}
\pagenumbering{arabic}

In the Standard Model~(SM) of particle physics, the $b\to s\gamma$ transition proceeds via loop Feynman diagrams.
The small size of the SM amplitude makes such process sensitive to the contribution of possible new particles.
The emitted photons are produced predominantly with left-handed helicity in the SM due to parity violation in the weak interaction, with a small relative right-handed component proportional to the ratio of $s$- to $b$-quark masses.
In many extensions of the SM, the right-handed component can be enhanced, leading to observable effects in mixing-induced \CP asymmetries and time-dependent decay rates of radiative \Bd and \Bs decays~\cite{Atwood97, Atwood05, Muheim08}.
Current measurements sensitive to right-handed contributions~\cite{Kspi0gamma-Belle, Kspi0gamma-BaBar, LHCb-PAPER-2012-019, LHCb-PAPER-2014-066, phigamma-Belle, LHCb-PAPER-2016-034} are in agreement with SM predictions~\cite{Paul17}.

The rate $\mathcal{P}(t)$ at which \Bs or \Bsb mesons decay to a common final state that contains a photon, such as $\phi\gamma$ (where $\phi$ refers to $\phi(1020)$), depends on the decay time $t$ as~\cite{Muheim08}
\begin{equation} \label{eq:decayrate}
\begin{split}
	\mathcal{P}(t) \propto e^{-\Gamma_s t} \bigl\{ &\cosh\left( \Delta\Gamma_s t / 2 \right) - \adeltavanilla\sinh\left( \Delta\Gamma_s t / 2 \right) \\ &+\zeta\,\Cvanilla\cos\left( \Delta m_s t \right) - \zeta\,\Svanilla\sin\left( \Delta m_s t \right) \bigr\},
\end{split}
\end{equation}
where $\Delta\Gamma_s$ and $\Delta m_s$ are the width and mass differences between the \Bs mass eigenstates, defined positively, $\Gamma_s$ is the mean decay width between such eigenstates, and $\zeta$ takes the value of $+1$ ($-1$) for an initial \Bs (\Bsb) state.
The coefficients \adeltavanilla and \Svanilla are sensitive to the photon helicity amplitudes and weak phases, while \Cvanilla is related to \CP violation in the decay.
The SM predictions for the three coefficients in the \phigamma decay are close to zero~\cite{Muheim08}.
The \lhcb collaboration has previously measured $\adelta = -0.98 \, ^{+0.46}_{-0.52} \, ^{+0.23} _{-0.20}$~\cite{LHCb-PAPER-2016-034} from a time-dependent flavour-untagged analysis, which is compatible with the SM within two standard deviations.

This Letter reports the first measurement of the \CP-violating observables \Svanilla and \Cvanilla from a radiative \Bs decay, determined from the time-dependent rate of \phigamma decays in which the $\phi$ meson decays to a $\Kp\Km$ pair. An update of the \adelta coefficient measurement is also provided.
Results are based on data collected with the LHCb detector in $pp$ collisions at center-of-mass energies of 7 and 8\tev during the years 2011 and 2012, respectively, corresponding to an integrated luminosity of 3\invfb.
Compared to Ref.~\cite{LHCb-PAPER-2016-034}, the current analysis benefits from a 20\% higher event selection efficiency, a reoptimized calorimeter reconstruction and a new photon identification algorithm.
Flavor-tagging algorithms are applied to determine the initial flavor of the \Bs or \Bsb meson, which is essential to measure the $S$ and $C$ observables, whereas flavor-untagged decays still contribute to the measurement of $\adeltavanilla$.
The background is subtracted from a fit to the mass distribution of the \Bs candidates.
A sample of untagged $\Bd \to \Kstarz\gamma$ decays (where \Kstarz refers to $\Kstarz(892)$), reconstructed in the flavor-specific $\Kstarz \to \Kp\pim$ final state, is used to control the decay-time-dependent efficiency, since its lifetime is well measured.
Throughout this Letter, the inclusion of charge-conjugated processes is implied.

The \lhcb detector is a single-arm forward spectrometer covering the pseudorapidity range 
$2 < \eta < 5$, described in detail in Refs.~\cite{Alves:2008zz,LHCb-DP-2014-002}.
It includes a high-precision tracking system consisting of a silicon-strip vertex detector surrounding the $pp$ interaction region, a large-area silicon-strip detector located upstream of a dipole magnet with a bending power of about $4{\mathrm{\,Tm}}$, and three stations of silicon-strip detectors and straw drift tubes placed downstream of the magnet.
Different types of charged hadrons are distinguished using information from two ring-imaging Cherenkov detectors. Photons, electrons and hadrons are identified by a calorimeter system consisting of scintillating-pad and preshower detectors, an electromagnetic and a hadronic calorimeter.

The online event selection is performed by a trigger system, which consists of a hardware stage, based on information from the calorimeter and muon systems, followed by a software stage, which applies a full event reconstruction.
Two trigger selections are defined, with different photon and track momentum thresholds.
Samples of simulated events, produced with the software described in Refs.~\cite{Sjostrand:2007gs, *Sjostrand:2006za, LHCb-PROC-2010-056, Lange:2001uf, Golonka:2005pn, Allison:2006ve, *Agostinelli:2002hh, LHCb-PROC-2011-006}, are used to characterize signal and background contributions.
The signal sample is generated with the three coefficients \adelta, \Cpar and \Spar set to zero.

Candidate \phigamma decays are reconstructed from a photon candidate and two oppositely charged particles identified as kaons.
The selection is designed to maximize the significance $S/\sqrt{S+B}$ of the signal yield.
Photons are reconstructed from energy deposits in the electromagnetic calorimeter and required to have a momentum transverse to the beam axis, \pt, larger than $3.0$ or $4.2 \gevc$, depending on the trigger selection.
Background due to photons from \piz decays is rejected by a dedicated algorithm~\cite{LHCb-PUB-2015-016}.
The kaon candidates are required to have $\ptot > 1.0\gevc$ and $\pt > 0.3\gevc$, where $\ptot$ is the total momentum, and at least one of them must fulfill $\ptot > 10 \gevc$ and $\pt > 1.2$ or $1.8\gevc$, depending on the trigger selection.
Kaon candidates are required to be inconsistent with originating from a primary $pp$ interaction vertex, and must form a common vertex of good quality.
The \Kp\Km system must have an invariant mass within $15 \mevcc$ of the known $\phi$ mass~\cite{PDG2018}.
The \Bs candidate must be consistent with originating from only one $pp$ interaction vertex, and only candidates with decay times between 0.3 and 10\ps are retained.
In addition, the cosine of the helicity angle ($\theta_H$), defined as the angle between the momenta of the positively charged kaon and that of the \Bs meson in the rest frame of the $\phi$ meson, is required to be less than $0.8$ in absolute value.
This requirement helps to suppress the \piz and combinatorial backgrounds, which are expected to be distributed as $\cos^2 \theta_H$ and a uniform distribution, respectively, as opposed to the $\sin^2 \theta_H$ distribution expected for the signal.
The $\Bd\to\Kstarz\gamma$ decay, with $\Kstarz \to \Kp\pim$, is selected with almost identical requirements. A pion is required instead of a kaon, and the invariant mass of the \Kp\pim system must be within $100 \mevcc$ of the known $\Kstarz$ mass~\cite{PDG2018}.

\begin{figure}[tb]
\centering
\includegraphics[width=0.49\textwidth]{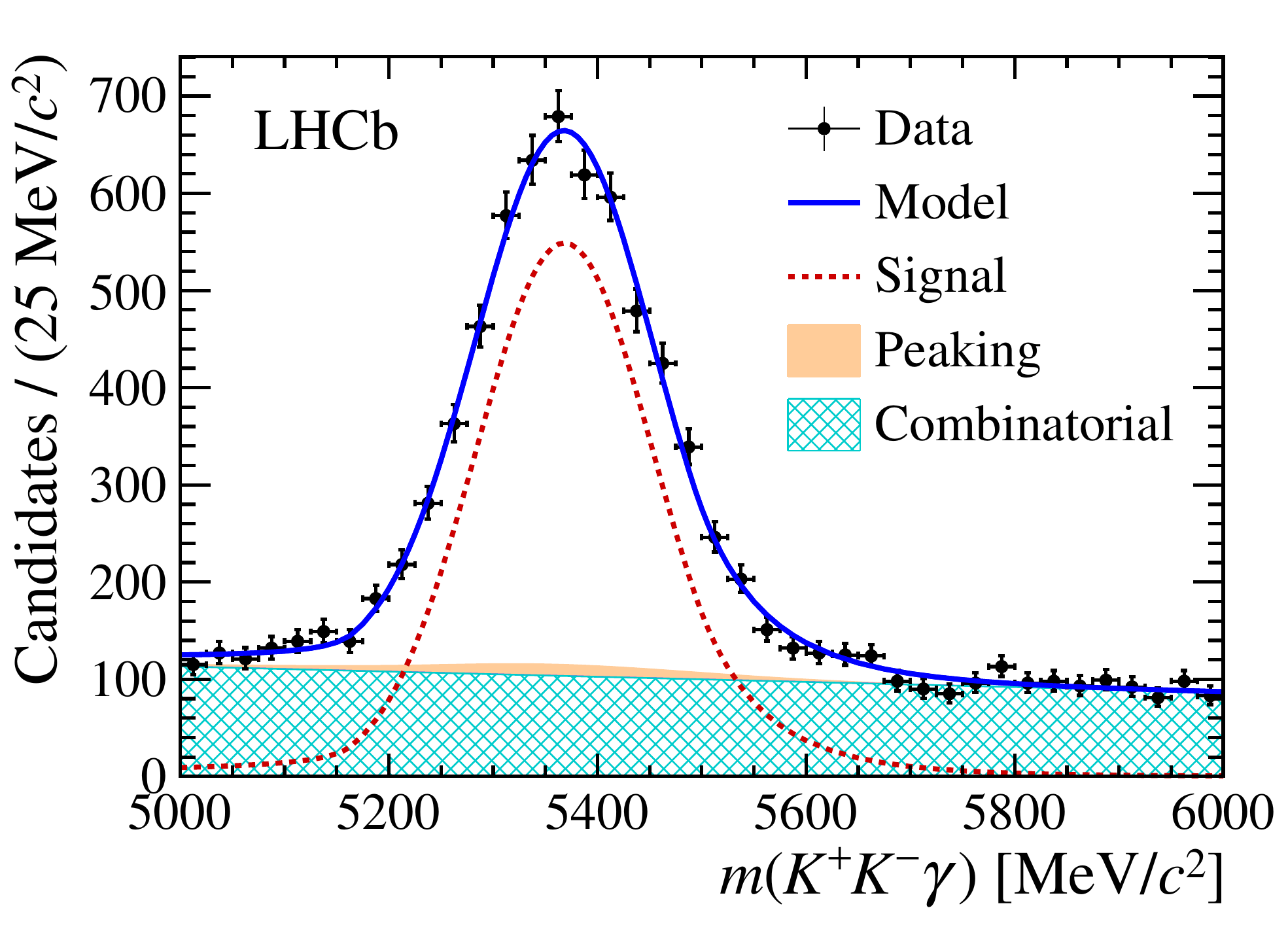}
\includegraphics[width=0.49\textwidth]{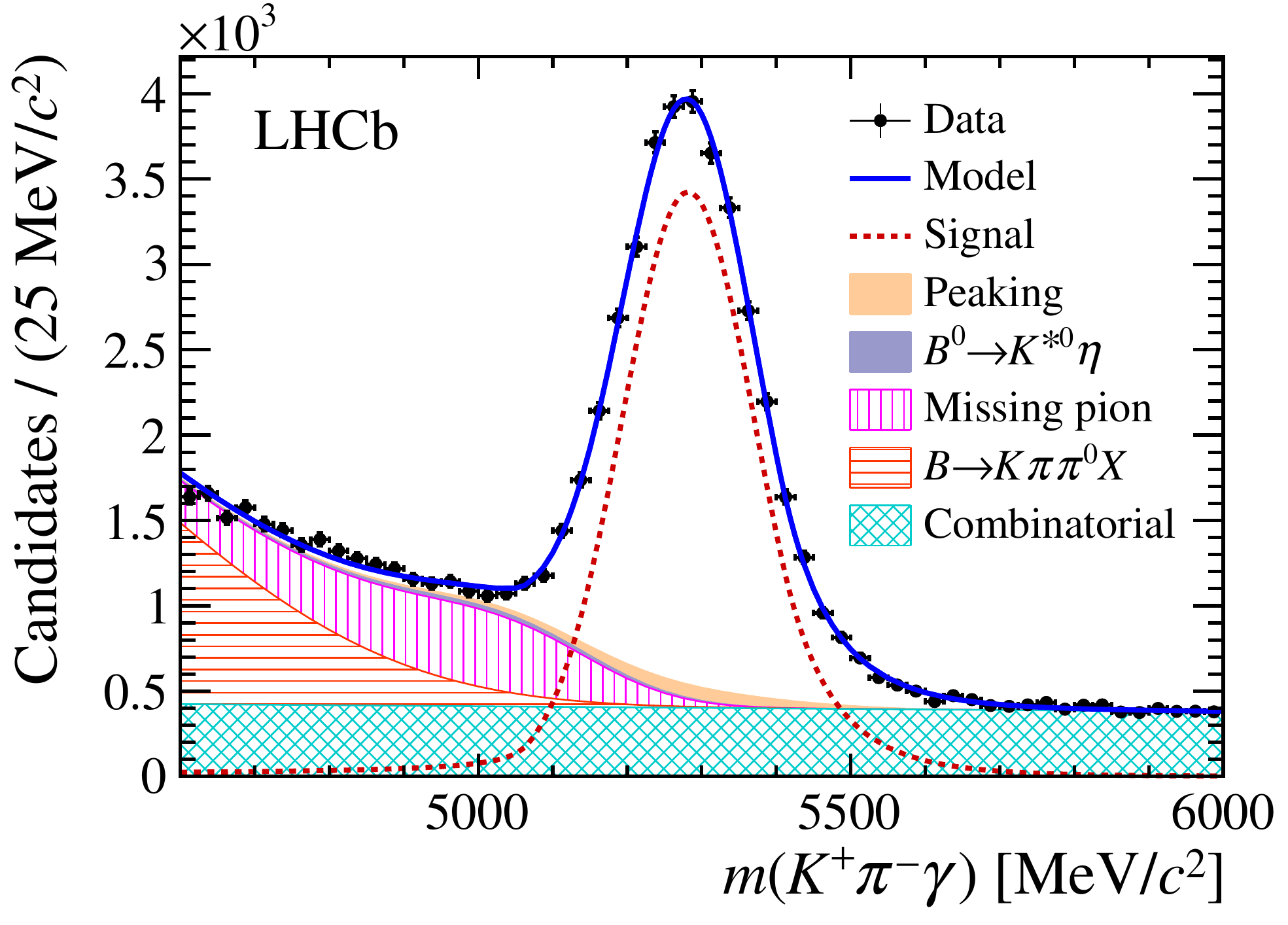}
\caption{\small Fits to the mass distributions
of the (left) \phigamma and (right) \kstgamma candidates.}
\label{fig:massfit}
\end{figure}

The signal yields are $5110 \pm 90$ for \phigamma decays and $33\,860 \pm 250$ for \kstgamma decays, where the uncertainties are statistical only. They are obtained from separate extended unbinned maximum-likelihood fits to the $\Bs\to\phi\gamma$ and $\Bd\to\Kstarz\gamma$ reconstructed mass distributions in the ranges 5000--6000\mevcc and 4600--6000\mevcc, respectively.
The mass fits are shown in Fig.~\ref{fig:massfit}.
The results are used to assign weights to the candidates in the data samples in order to subtract the backgrounds~\cite{Pivk:2004ty}.
The signal line shapes are described by modified Crystal Ball functions~\cite{Skwarnicki:1986xj}, consisting of a Gaussian core with power-law tails on both sides of the peak.
The mean and width of the Gaussian core are obtained from data, while the tail parameters are determined from simulation.
Three background categories are considered: combinatorial, peaking, and partially reconstructed.
The combinatorial background, modeled by a linear function, is produced by the wrong association of a random photon with two hadrons mostly coming from real $\phi$ and $\Kstarz$ resonances.
The peaking backgrounds originate from other $b$-hadron decays with a reconstructed mass falling under the signal peak, due to the misidentification of one or several final-state particles.
All possible combinations of misidentified hadrons, or the misidentification of a $\pi^0$ meson as a photon, are considered for the signal and control decay channels.
For the \phigamma decay channel, the relevant contributions are $\Bs \to \phi\pi^0$ and $\Lb \to (p\Km)\gamma$, where $p\Km$ comes from $\Lambda(1520)$ and further baryon resonances. For the \kstgamma decay channel, the $\Bd \to \Kstarz\pi^0$ and $\Lb \to (p\Km)\gamma$ decays are taken into account.
Each peaking background is modeled with a Crystal Ball function. The shape parameters are determined from simulation, except for the width of the Gaussian core, which is multiplied by a factor to account for the difference in resolution between data and simulation.
The yield ratios of peaking backgrounds to signal are calculated using simulation samples and taking the branching ratios from experimental measurements~\cite{LHCb-PAPER-2016-034, LHCb-PAPER-2012-019}.
They are determined to be below 2\% in all cases.
Partially reconstructed backgrounds originate from other $b$-hadron decays in which one or several final-state particles are not reconstructed.
This contribution is negligible in \phigamma decays, while for the \kstgamma mode the dominant contributions are: decays of the type $B \to K\pi\pi\gamma$ with a missing pion, decays of the type $B \to K\pi\pi^0 X$ (mainly from $B^+ \to \Dzb\rho^+$ decays) with one or several missing hadrons, and $\Bd \to \Kstarz\eta(\gamma\gamma)$ decays with a missing photon.
They are described by an ARGUS function~\cite{ARGUS-function} convolved with a Gaussian function to account for the detector resolution, with the shape parameters determined from simulation.

Flavor-tagging algorithms are applied to identify the initial flavor of the $\Bs$ meson. They provide a tag decision $q$, which takes the value $+1$ if the signal was originally a \Bs meson, $-1$ if it was a \Bsb meson, and zero if no decision is given. The algorithms also provide an estimate $\eta$ of the probability for the tag decision to be incorrect (mistag probability).
Two classes of flavor-tagging algorithms are used: same-side (SS)~\cite{LHCb-PAPER-2015-056} and opposite-side (OS) taggers~\cite{LHCb-PAPER-2011-027}.
The SS tagger determines the flavor of the signal candidate by identifying the charge of the kaon produced together with the \Bs meson in the fragmentation process, and is based on a neural network algorithm~\cite{LHCb-PAPER-2015-056}.
The OS taggers rely on the pair production of $b$ hadrons in $pp$ collisions and examine the decay products of the other $b$ hadron in the event.
The information used includes the charge of the leptons produced in semileptonic decays, the charge of kaons produced in $b \to c \to s$ transitions, and the charge of the particles originating from the decay vertex~\cite{LHCb-PAPER-2011-027}.

The mistag probability estimate $\eta$ is calibrated using a linear function to obtain a corrected mistag probability $\omega$ for the signal sample.
This is performed using mainly samples of $B^+ \to \jpsi \Kp$ and $B^0 \to \jpsi\Kstarz$ decays for the OS tagger and $\Bs \to D_s^-\pi^+$ and $B^*_{s2}(5840)^0 \to B^+ \Km$ decays for the SS tagger.
The uncertainties of the calibration parameters include a systematic uncertainty that takes into account possible differences of these parameters between the decays used for calibration and other $B$-decay modes. The validity of these calibrations for \phigamma decays is checked using both simulation and data.
Finally, the outputs of the algorithms are combined into a single decision and mistag probability.
The effective tagging efficiency, $\epsilon_{\text{eff}} = \left( 4.99 \pm 0.14 \right)$\%, is the product of the probability to obtain a decision $\epsilon_{\text{tag}} =  \left( 74.5 \pm 0.8 \right)$\% and the square of the effective dilution $D = 1 - 2\omega = \left( 25.9 \pm 0.3 \right)$\%.

The \CP-violating and mixing-induced observables are determined from a weighted unbinned maximum-likelihood fit~\cite{Xie:2009rka} to the decay-time distributions, performed simultaneously on the \phigamma and \kstgamma samples.
The signal probability density function (\PDF) of the \phigamma decay-time distribution is defined as the decay rate $\mathcal{P}(t)$ in Eq. \ref{eq:decayrate}, convolved with a resolution function and multiplied by a decay-time-dependent efficiency $\epsilon(t)$.
For the \kstgamma decay, the time-dependent decay rate is described as a single exponential function.
The physics parameters are constrained to the averages from Ref.~\cite{HFLAV16}: $\tau_{\Bd} = 1.520 \pm 0.004\ps$, $\Gamma_{s} = 0.6629 \pm 0.0018\invps$, $\Delta\Gamma_s = 0.088 \pm 0.006\invps$ and $\Delta m_s = 17.757 \pm 0.021\invps$. The correlation of $-0.11$ between the $\Gamma_s$ and $\Delta\Gamma_s$ parameters is taken into account.

The decay-time resolution is modeled by the sum of two Gaussian functions, with a common mean and independent widths.
The widths are given by the per-candidate decay-time uncertainties, multiplied by constant scaling factors determined from simulation to account for an observed underestimation of the uncertainties.
Additional control samples are used to determine the decay-time resolution differences between simulation and data, which are accounted for in the analysis as a source of systematic uncertainty.
These samples include $\phi$ mesons coming from $pp$ interaction vertices and $\Bd \to \jpsi\Kstarz$ decays, with $\jpsi \to \mup\mun$. In the latter case, in order to emulate the signal behavior, the decay is reconstructed with the two muons not contributing to the vertex fitting.
The resolution depends strongly on the decay time, with an average of $70\fs$.
The decay-time resolution is dominated by the photon momentum resolution, therefore being similar for \phigamma and \kstgamma decays.

The efficiency as a function of the decay time $t$ is parameterized as
\begin{equation}
    \epsilon(t) \propto \dfrac{t^{a/t}}{\cosh(b\,t)} \,,
\end{equation}
where the parameters $a$ and $b$ describe mainly the shape of the function at low and high decay times, respectively.
One hundred bins of variable size are defined to characterize this function.
The efficiency parameters are determined in the simultaneous fit to the data, mainly driven from \kstgamma candidates, while the differences between the two decays are obtained from simulation and fixed in the data fit.
In simulation, the decay-time-dependent efficiencies of the two decay modes are compatible within uncertainties.

Pseudoexperiments are used to validate the overall fit procedure. In each pseudoexperiment, samples of \phigamma and \kstgamma signal decays are generated based on the data mass fit and the expected yields. Background candidates are included taking random events from data or simulation.
The mass and the decay-time fits are then performed, following the nominal procedure.
The procedure is repeated for several values of the coefficients.
No biases are found on the average fitted values, in any scenario. Statistical uncertainties are found to be underestimated by about 15\% for \Spar and \Cpar, and 5\% for \adelta, due to the background-subtraction weights~\cite{Xie:2009rka}. The uncertainties are corrected for in the results below.

\begin{figure}[tb]
\centering
\includegraphics[width=0.49\textwidth]{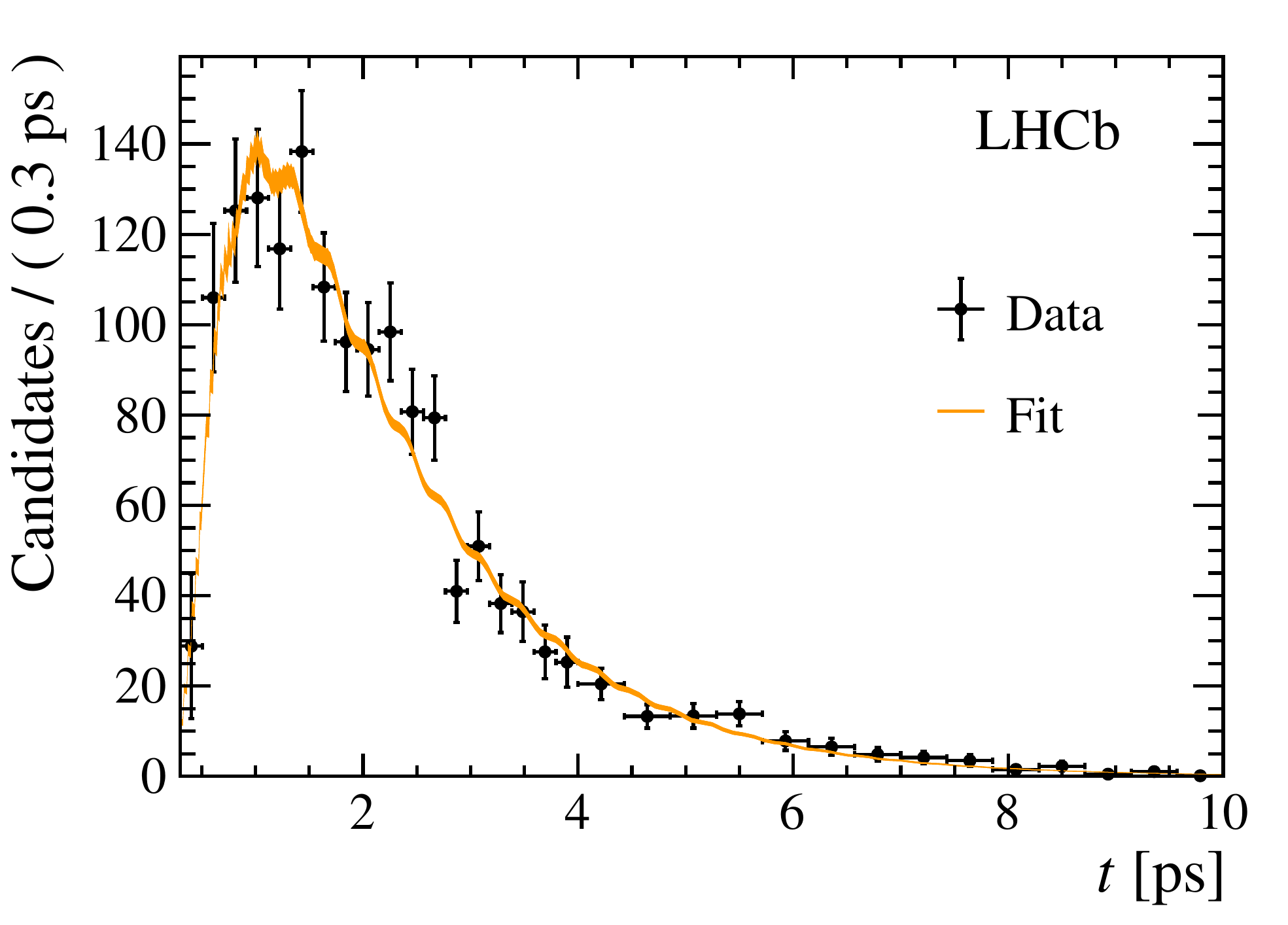}
\includegraphics[width=0.49\textwidth]{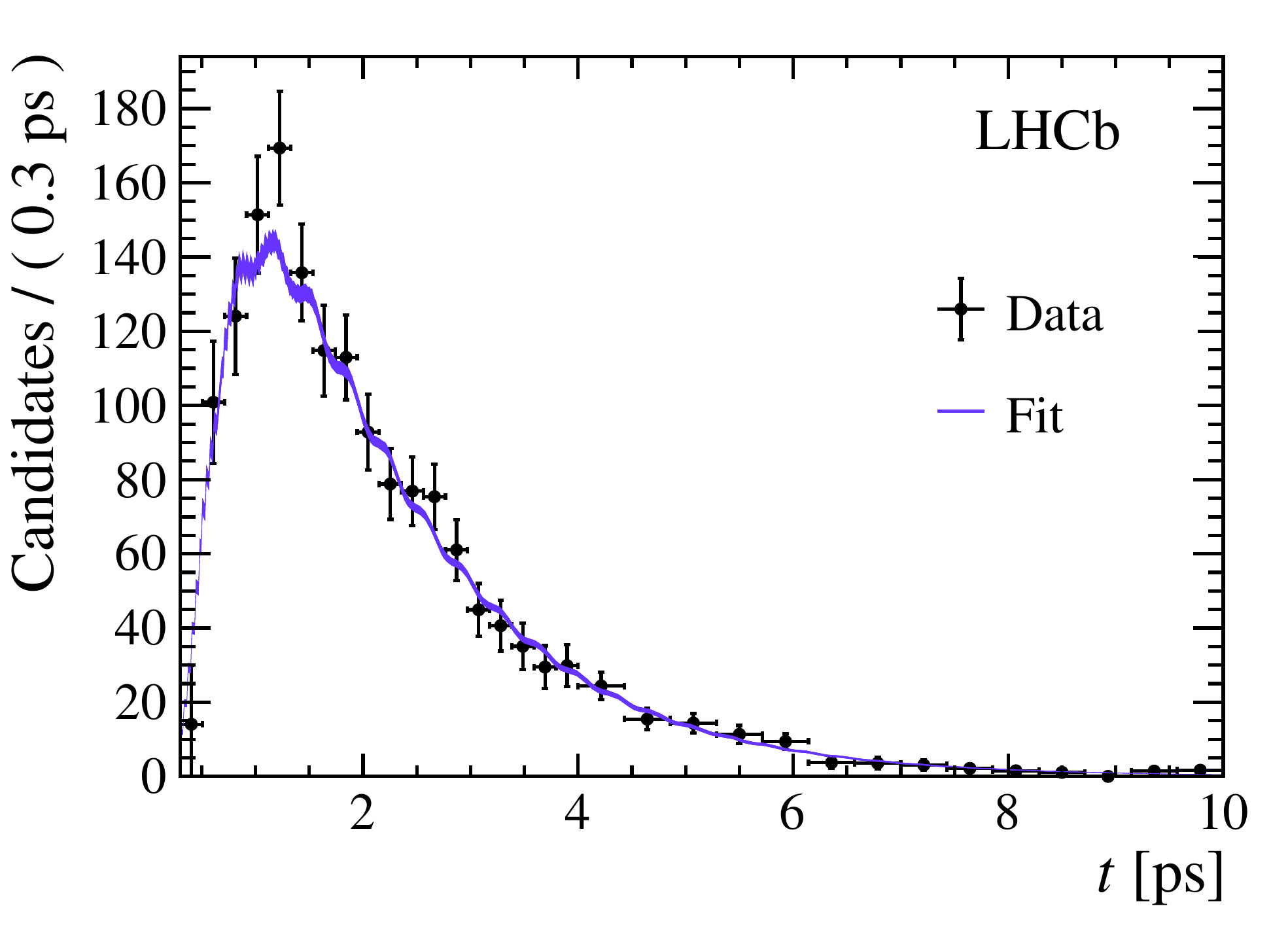} \\
\includegraphics[width=0.49\textwidth]{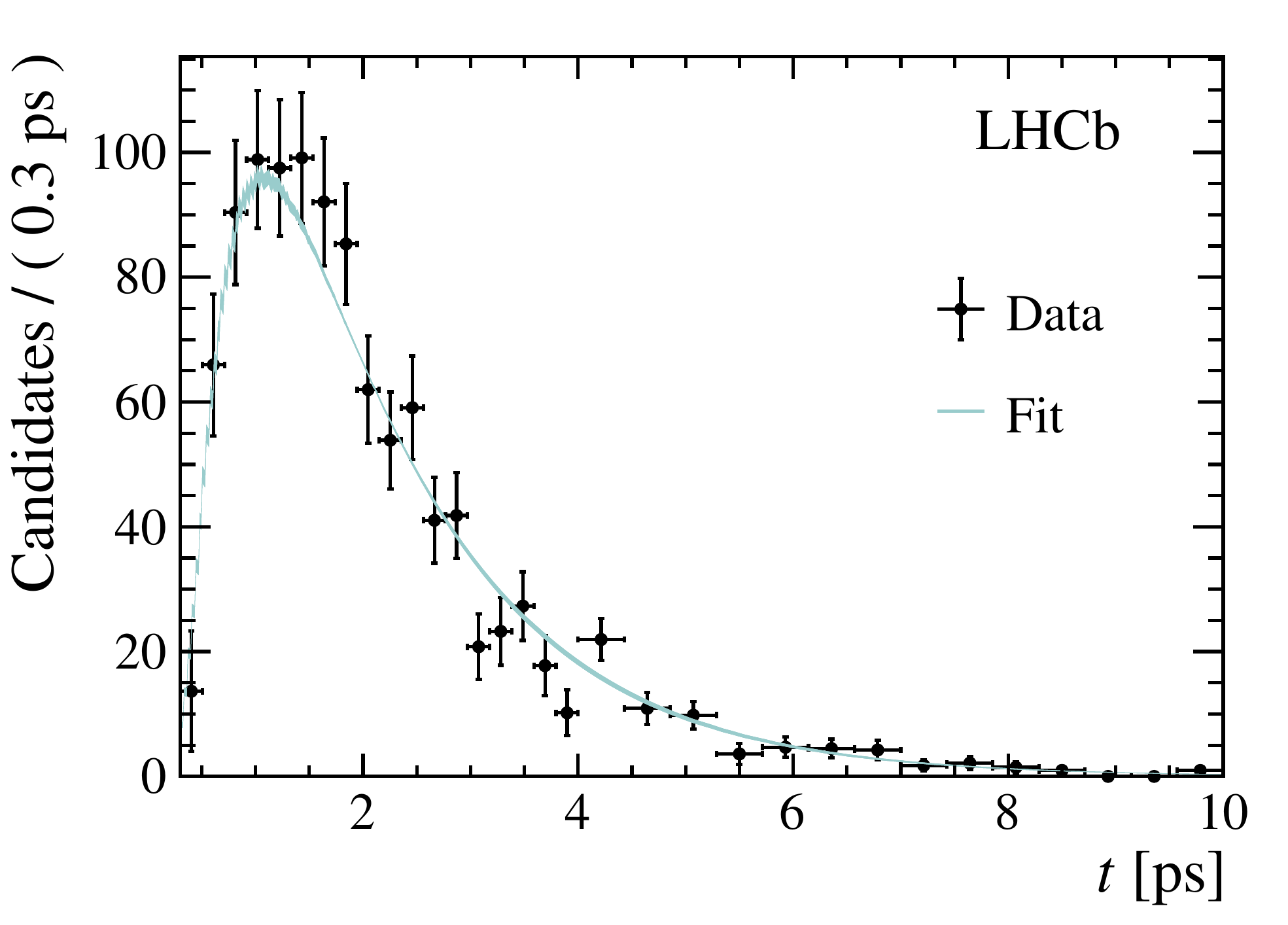}
\includegraphics[width=0.49\textwidth]{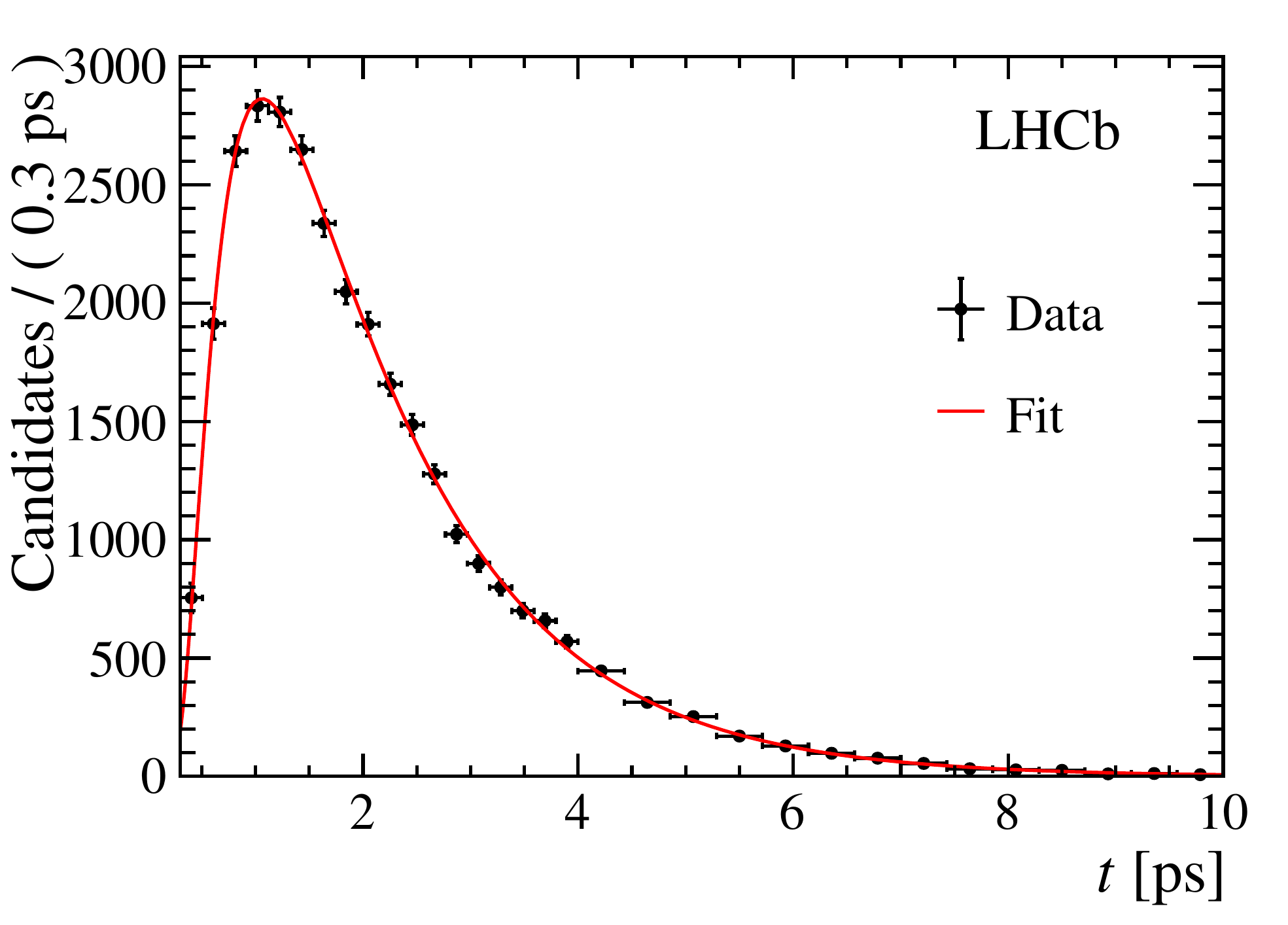}
\caption{\small
Decay-time fit projections.
The top row corresponds to the tagged (left) $\phigamma$ and (right) $\Bsb \to \phi\gamma$ candidates, while the bottom plots show the (left) untagged $\phigamma$ and (right) $\kstgamma$ candidates.
The line is the result of the fit described in the text, including statistical uncertainties.
}
\label{fig:decaytimefit}
\end{figure}

The decay-time distributions and the corresponding fit projections are shown in Fig.~\ref{fig:decaytimefit}. The fitted values are $\Spar = 0.43 \pm 0.30$, $\Cpar = 0.11 \pm 0.29$ and $\adelta = -0.67 \, ^{+0.37}_{-0.41}$, with a small correlation of $-0.04$ between each pair of observables.
The statistical uncertainty includes the uncertainty from the physics parameters taken from external measurements.
For \Spar and \Cpar, the systematic uncertainty is dominated by the effects of possible differences between data and simulation in the decay-time resolution parameters (0.08), and the uncertainty on the parameters used to calibrate the same-side tagging algorithms (0.04).
For \adelta, the dominant source of systematic uncertainties is related to the determination of the decay-time-dependent efficiency function. In particular, the contribution of the partially reconstructed background of $\kstgamma$ decays, coming from the correlation between reconstructed mass and time (0.11) and the mass-shape modeling (0.08), and the limited size of the simulation sample used to determine the efficiency differences between \phigamma and \kstgamma decays (0.08).
The total systematic uncertainties are $0.11$ for \Spar and \Cpar, and $0.17$ for \adelta.

In summary, the \CP-violating and mixing-induced observables $\Spar$, $\Cpar$ and $\adelta$ are measured from a time-dependent analysis of \phigamma decays, using a data sample corresponding to an integrated luminosity of 3\invfb collected with the LHCb experiment during the 2011 and 2012 data-taking periods.
More than $5000$ \phigamma decays are reconstructed.
A sample of \kstgamma decays, which is six times larger, is used for the calibration of the time-dependent efficiency.
From a simultaneous unbinned fit to the \phigamma and \kstgamma data samples, the values
\begin{align*}
  	\Spar &= 0.43 \pm 0.30 \pm 0.11, \\
  	\Cpar &= 0.11 \pm 0.29 \pm 0.11, \\
  	\adelta &= -0.67 \, ^{+0.37}_{-0.41} \pm 0.17
\end{align*}
are measured, where the first uncertainty is statistical and the second systematic. The results are compatible with the SM expectation~\cite{Muheim08} within $1.3$, $0.3$ and $1.7$ standard deviations, respectively.

\section*{Acknowledgements}
%
%
\noindent We express our gratitude to our colleagues in the CERN
accelerator departments for the excellent performance of the LHC. We
thank the technical and administrative staff at the LHCb
institutes.
We acknowledge support from CERN and from the national agencies:
CAPES, CNPq, FAPERJ and FINEP (Brazil); 
MOST and NSFC (China); 
CNRS/IN2P3 (France); 
BMBF, DFG and MPG (Germany); 
INFN (Italy); 
NWO (Netherlands); 
MNiSW and NCN (Poland); 
MEN/IFA (Romania); 
MSHE (Russia); 
MinECo (Spain); 
SNSF and SER (Switzerland); 
NASU (Ukraine); 
STFC (United Kingdom); 
DOE NP and NSF (USA).
We acknowledge the computing resources that are provided by CERN, IN2P3
(France), KIT and DESY (Germany), INFN (Italy), SURF (Netherlands),
PIC (Spain), GridPP (United Kingdom), RRCKI and Yandex
LLC (Russia), CSCS (Switzerland), IFIN-HH (Romania), CBPF (Brazil),
PL-GRID (Poland) and OSC (USA).
We are indebted to the communities behind the multiple open-source
software packages on which we depend.
Individual groups or members have received support from
AvH Foundation (Germany);
EPLANET, Marie Sk\l{}odowska-Curie Actions and ERC (European Union);
ANR, Labex P2IO and OCEVU, and R\'{e}gion Auvergne-Rh\^{o}ne-Alpes (France);
Key Research Program of Frontier Sciences of CAS, CAS PIFI, and the Thousand Talents Program (China);
RFBR, RSF and Yandex LLC (Russia);
GVA, XuntaGal and GENCAT (Spain);
the Royal Society
and the Leverhulme Trust (United Kingdom).

\appendix

\addcontentsline{toc}{section}{References}
\bibliographystyle{LHCb}
\bibliography{ms}

\clearpage
\centerline
{\large\bf LHCb collaboration}
\begin
{flushleft}
\small
R.~Aaij$^{29}$,
C.~Abell{\'a}n~Beteta$^{46}$,
B.~Adeva$^{43}$,
M.~Adinolfi$^{50}$,
C.A.~Aidala$^{77}$,
Z.~Ajaltouni$^{7}$,
S.~Akar$^{61}$,
P.~Albicocco$^{20}$,
J.~Albrecht$^{12}$,
F.~Alessio$^{44}$
% ...

{\footnotesize \it

% ...
$ ^{7}$Universit{\'e} Clermont Auvergne, CNRS/IN2P3, LPC, Clermont-Ferrand, France\\
% ...
$ ^{12}$Fakult{\"a}t Physik, Technische Universit{\"a}t Dortmund, Dortmund, Germany\\
% ...
$ ^{20}$INFN Laboratori Nazionali di Frascati, Frascati, Italy\\
% ...
$ ^{29}$Nikhef National Institute for Subatomic Physics, Amsterdam, Netherlands\\
% ...
$ ^{43}$Instituto Galego de F{\'\i}sica de Altas Enerx{\'\i}as (IGFAE), Universidade de Santiago de Compostela, Santiago de Compostela, Spain\\
$ ^{44}$European Organization for Nuclear Research (CERN), Geneva, Switzerland\\
% ...
$ ^{46}$Physik-Institut, Universit{\"a}t Z{\"u}rich, Z{\"u}rich, Switzerland\\
% ...
$ ^{50}$H.H. Wills Physics Laboratory, University of Bristol, Bristol, United Kingdom\\
% ...
$ ^{61}$University of Cincinnati, Cincinnati, OH, United States\\
% ...
$ ^{63}$Syracuse University, Syracuse, NY, United States\\
% ...
$ ^{77}$University of Michigan, Ann Arbor, United States, associated to $^{63}$\\
% ...
}
\end{flushleft}

\end{document}